# Tracing Noble Gas Radionuclides in the Environment

Philippe Collon

*Physics Department, University of Notre Dame, Notre Dame, Indiana 46556;*
*e-mail: pcollon@nd.edu*

Walter Kutschera

*Institut für Isotopenforschung und Kernphysik, Universität Wien, A-1090 Wien, Austria;*
*e-mail: walter.kutschera@univie.ac.at*

Zheng-Tian Lu

*Physics Division, Argonne National Laboratory, Argonne, Illinois 60439;*
*e-mail: lu@anl.gov*

**Key Words**

$^{39}$Ar, $^{81}$Kr, $^{85}$Kr, Low Level Counting, Accelerator Mass Spectrometry, Atom Trap Analysis

**Abstract**

   Trace analysis of radionuclides is an essential and versatile tool in modern science and technology. Due to their ideal geophysical and geochemical properties, long-lived noble gas radionuclides, in particular, $^{39}$Ar ($t_{1/2} = 269$ yr), $^{81}$Kr ($t_{1/2} = 2.3 \times 10^5$ yr) and $^{85}$Kr ($t_{1/2} = 10.8$ yr), have long been recognized to have a wide range of important applications in Earth sciences. In recent years, significant progress has been made in the development of practical analytical methods, and has led to applications of these isotopes in the hydrosphere (tracing the flow of groundwater and ocean water). In this article, we introduce the applications of these isotopes and review three leading analytical methods: Low-Level Counting (LLC), Accelerator Mass Spectrometry (AMS) and Atom Trap Trace Analysis (ATTA).

**CONTENTS**

# 1. INTRODUCTION

We are becoming increasingly aware that natural resources on Earth are limited and cannot be used in an uncontrolled way without irreversible consequences. Moreover, it is now scientifically established that human activities are affecting the environment in which we live. However, in trying to address these and other questions we are faced with one major problem: We lack essential data that would enable us to understand more completely how our environment is being affected. At present it is often not possible to decide unequivocally whether these observed changes are natural or anthropogenic.

Despite enormous progress made in environmental science in the past few decades, we still have inadequate knowledge as to how different systems of our environment are interconnected and how human activities are changing these systems. Such knowledge is essential to the extremely challenging task of developing realistic models of the environment, which is not a controlled system in a laboratory but one that is highly complex. Not only does the system change from one geographic location to the next, but it also changes from one period to the next. In order to interpret both the short-term and long-term effects of various emissions into our environment, we need a better understanding of the various transport mechanisms and mixing processes that take place in our environment, and we need better tools to probe theses processes. One such powerful tool is the use of both natural and anthropogenic nuclides as environmental tracers. They are used to trace the main transport and mixing processes in order to help determine the dynamics of the environmental system and the influence of human activity.

## 1.1. Origin and classification of radionuclides

Radioactive nuclides, or radionuclides, have possible natural and anthropogenic origins. Natural radionuclides can be further classified, according to their origins, into four major categories:

1) *Stellar* (e.g. $^{235,\,238}$U, $^{232}$Th, $^{87}$Rb, $^{40}$K). Their half-lives are usually comparable to the age of the solar system ($10^9$ yr). They were produced in stellar nucleosynthesis (1) and were already present in the solar system during the formation of Earth. Remnant concentrations of these nuclides can still be detected. For example, uranium has a concentration of ~ 2.2 ppm in the Earth's crust, which corresponds to about half the amount present at the formation of the Earth 4.5 billion years ago.

2) *Radiogenic* (e.g. $^{222}$Rn, $^{226}$Ra). They are decay products of natural decay series of uranium and thorium.

3) *Cosmogenic* (e.g. $^{14}$C, $^{39}$Ar, $^{41}$Ca, $^{81}$Kr). They are produced by cosmic-ray induced reactions in the atmosphere and by reactions due to secondary cosmic-rays (neutrons and muons) at the surface of the lithosphere.

4) *Fissiogenic* (e.g. $^{85}$Kr, $^{90}$Sr, $^{129}$I). They are produced through fission processes (spontaneous and neutron induced) occurring in the lithosphere. The abundances of these isotopes in the environment have been greatly affected by man-made nuclear fissions.

Anthropogenic radionuclides encompass all radioactive nuclides that find their origin in human activities including nuclear industry, atomic bomb tests, medical applications, etc. Some of these anthropogenic radionuclides do not naturally occur in our environment and can therefore be used to monitor human activities as well as natural processes that have taken place in the past five decades since the dawn of the nuclear age.

Among these categories, those cosmogenic radionuclides produced in the atmosphere are of considerable interest in environmental science. Their production, transport, and deposition processes have been extensively studied (2, 3). The introduction of these radionuclides into the atmosphere occurs continuously at a rate that, in principle, can be determined both in space and time. Cosmic-ray induced nuclear reactions in the atmosphere are simulated with computer codes based on atmospheric models and nuclear data in order to establish accurate description of the production rates and the distribution functions of various cosmogenic radionuclides. The produced radionuclides subsequently go through mixing processes in the atmosphere. These are governed by their physical and chemical properties. Some of the radionuclides are removed from the atmosphere by either precipitation or exchange processes at the ground level. In this way they enter different archives on Earth including the atmosphere, biosphere, hydrosphere, cryosphere, and lithosphere (4). Measurements of isotopic concentration in these archives provide information about the source function and effective introduction of these radionuclides into a specific environment. This information, then, forms the basis of using these isotope tracers to study environmental processes.

## 1.2. Noble gases

Among those cosmogenic radionuclides originating in the atmosphere, noble gas radionuclides could play a particularly important role as tracers in environmental studies due to their unique chemical inertness. There are three noble gas radionuclides in nature with half-lives that are sufficiently long to allow geological applications: $^{39}$Ar ($t_{1/2}$ = 269 yr), $^{81}$Kr ($t_{1/2}$ = 2.3×10$^5$ yr) and $^{85}$Kr ($t_{1/2}$ = 10.8 yr) (5). $^{39}$Ar and $^{81}$Kr in the atmosphere are mainly cosmogenic in origin, while $^{85}$Kr is dominated by anthropogenic fission products.

Due to their chemical inertness and low solubilities in water, almost all krypton and argon gases reside in the atmosphere. The concentration of [39]Ar and [81]Kr is spatially uniform around the globe due to their long residence times and extensive mixing processes. [85]Kr is less so, due to its shorter half-life. In practice, the chemical inertness of noble gas presents the analysts both an advantage and a disadvantage. On one hand, noble gas can be extracted from a large quantity (tons) of environmental samples (water, ice) with high efficiency (> 50%); on the other hand, the low solubilities of noble gases result in low concentration in the environmental samples in the first place.

The applicable age range of a radionuclide closely follows its half-life, and is affected by both the measurement uncertainties and the knowledge of its initial abundance. A simplified rule of thumb is that the applicable age ranges from 0.1–10 times the half-life. Radionuclides of different half-lives are used to study effects of different age ranges, as well as to unravel the different mixing and diffusion processes in a specific environment. They complement each other, and their overlapping age ranges ideally should link together to probe a continuous history of the environment. Fig. 1 shows the applicable age ranges of the three noble gas radionuclides along with those of a few other popular radionuclides.

## 2. PROPERTIES of [39]Ar, [81]Kr and [85]Kr

### 2.1. [39]Ar

[39]Ar was first discovered by Brosli et al. in irradiated potassium salts (6). It β-decays to the ground-level of [39]K with no γ-ray emissions (5). Its half-life (269±3 yr) (7) lies in between, and near the geometric mean of, the half-lives of two popular dating isotopes [3]H (12 yr) and [14]C (5715 yr). This favorable half-life, combined with the aforementioned advantages of a noble gas tracer, makes [39]Ar a desirable isotope for dating underground and ocean water samples in the age range of 50–1000 years.

In nature, [39]Ar is mainly produced by cosmic-ray induced [40]Ar(n,2n)[39]Ar reactions in the atmosphere (8). In tropospheric argon samples, Loosli measured the activity of [39]Ar to be 0.107±0.004 dpm per liter of Ar (all gas volumes are STP volumes throughout this paper) (9), which corresponds to an atmospheric ratio of [39]Ar/Ar = $(8.1\pm0.3)\times10^{-16}$. Variations of this ratio of up to 7% in the last 1000 years, due to the fluctuation in cosmic-ray flux, were estimated based on [14]C data in tree rings (10), and can be corrected for in [39]Ar-dating applications. Loosli compared the [39]Ar activities between pre-bomb and post-bomb (these terms, respectively, refer to periods before and after the dawn of the nuclear age around 1950) samples and found that the anthropogenic contribution to the atmospheric $^{39}$Ar inventory is less than 5% (8).

As a noble gas, $^{39}$Ar is not involved in chemical processes and its distribution in terrestrial reservoirs is relatively simple: 99% is concentrated in the atmosphere with an activity that is in quasi-steady state, and based on its solubility one estimates ~1% residing in the oceans. There can be substantial subsurface production in granite rocks through the $^{39}$K(n, p)$^{39}$Ar reaction (11, 12). This in-situ production source is likely to affect the reliability of $^{39}$Ar dating of groundwater, but is not a concern for dating ocean water.

## 2.2. $^{81}$Kr

$^{81}$Kr was first discovered by Reynolds in 1950 after intense neutron activation of a bromine target (13). It decays via electron capture with a 99.7% branch to the ground-level of $^{81}$Br, and a weak 0.3% branch to an excited level that promptly decays with a 276 keV γ-ray emission (5, 14). Following K-capture there is a 13.5 keV characteristic x-ray emission. By counting the x-ray emissions in an enriched $^{81}$Kr sample, its half-life was determined to be $(2.13\pm0.16)\times10^5$ years (15). However, a recent re-evaluation based on a new value of K-capture fluorescence yield led to a revised half-life of $(2.29\pm0.11)\times10^5$ years (16). This long half-life, 40 times of that of $^{14}$C, makes $^{81}$Kr a desirable isotope for dating in the age range $5\times10^4$–$2\times10^6$ yr. Although an extensive amount of work has been performed in this age range with two other long-lived isotopes $^{10}$Be ($1.5\times10^6$ yr) and $^{36}$Cl ($3.0\times10^5$ yr), the interpretation of the $^{10}$Be and $^{36}$Cl data have been hindered by the difficult task of modeling the complicated production, transport and deposition processes of these two isotopes (17, 18). In contrast, $^{81}$Kr is produced and resides in the atmosphere, and enjoys a spatially homogeneous and temporally constant initial abundance. It is the difficulty of measuring $^{81}$Kr/Kr ratios that has been the main drawback of $^{81}$Kr dating.

The main nuclear reactions leading to cosmogenic $^{81}$Kr are proton- and neutron-induced spallation of stable $^{82}$Kr, $^{83}$Kr, $^{84}$Kr and $^{86}$Kr, as well as nuclear reactions $^{80}$Kr(n,γ)$^{81}$Kr and $^{82}$Kr(γ,n)$^{81}$Kr (19). Three independent measurements gave an atmospheric $^{81}$Kr/Kr ratio (unweighted mean) of $(5.2\pm0.4)\times10^{-13}$ (9, 20, 21). Using this value and assuming a secular equilibrium in the atmosphere we can determine the atmospheric production rate to be $(1.2\pm0.1)\times10^{-6}$ $^{81}$Kr atoms/cm$^2$/s, in good agreement with the theoretical value of $(1.1\pm0.2)\times10^{-6}$ atoms/cm$^2$/s calculated by J. Mazarik (22).

The atmosphere can be considered as the only major reservoir of $^{81}$Kr on Earth, similar to $^{39}$Ar. Indeed 97.5% of $^{81}$Kr is in the atmosphere, and based on the solubility of krypton in water one estimates that 2.5% resides in the ocean. This is in strong contrast to [14]C: 93% of the global [14]C inventory is stored in the ocean, while only 2% residing in the atmosphere and 5 % in the biosphere (23). It is easy to imagine that exchanges between the ocean and the atmosphere have a large effect on the atmospheric [14]C content, whereas very little influence is expected by such exchanges on the atmospheric [81]Kr inventory.

Fission products of [238]U are rich in neutrons and in general decay towards the stable isotopes via successive β$^-$ emissions. [81]Kr is shielded by the stable [81]Br from the neutron-rich nuclei and is therefore not populated via the isobaric decay chain. The direct fission yield of [81]Kr is estimated to be only $7\times10^{-11}$ (11) and results in an equilibrium concentration of $5\times10^{-3}$ [81]Kr atoms per ton of rock with 1 ppm content of uranium. Compared to atmospheric production, contributions from both spontaneous and neutron-induced fission and from human activities are negligible (24).

## 2.3. [85]Kr

[85]Kr was discovered among the products of uranium fission. Its radioactivity was first published by Zeldes et al (25). It β-decays with a 99.56% branch to the ground-level of [85]Rb and a weak 0.43% branch to an excited level followed by a 514 keV γ-ray emission (5). It has a half-life of 10.76±0.02 years (26).

With no stable isotopes shielding it from the neutron-rich side, [85]Kr is amply produced in uranium and plutonium fission. While it is produced naturally, anthropogenic production of [85]Kr, from fission of uranium and plutonium in nuclear reactors and released into the atmosphere due to reprocessing of spent nuclear fuel rods (27), has resulted in a dramatic increase of [85]Kr by a factor of $10^6$ in the atmosphere since the early 1950s (Fig. 2). The atmospheric [85]Kr activity is approximately 1 Bq/m$^3$ air, which corresponds to an atmospheric ratio of [85]Kr/Kr $\approx 2\times10^{-11}$. Due to its relatively short half-life, the spatial distribution of [85]Kr is not as constant as [39]Ar and [81]Kr. For example, [85]Kr/Kr in the northern hemisphere, where most of nuclear fuel reprocessing plants reside, can be ~ 20% higher than that in the southern hemisphere (28). Due to frequent gas releases, [85]Kr/Kr near a nuclear fuel reprocessing plant can experience brief elevated levels above its long-term average value.

This increase in atmospheric concentration has been well documented and [85]Kr is a useful tracer of circulation and mixing in the atmosphere (29 – 31). [85]Kr enters natural water systems from the atmosphere and is very useful for dating recently formed groundwater (32 – 36) and subsurface water in the oceans (37, 38). The concentration of [85]Kr in the atmosphere is continuously monitored as a means to verify compliance to the Nuclear Non-Proliferation Treaty. In addition, it has been proposed that [85]Kr can serve as a leak tracer to monitor the mechanical integrity of nuclear fuel rods.

## 3. DETECTION TECHNIQUES

For all the reasons detailed above, it would be natural to ask the question why these tracers are not at present used on a regular basis in environmental science. The bottleneck that has so far prevented the widespread use of these three isotopes has been the lack of a convenient analytical method. Part of the answer lies in the low isotopic ratios of these radionuclides, which combined with their respective solubilities, results in extremely low concentrations of these tracers in water samples. To illustrate this point let us consider one liter of typical "modern" ocean water, which contains 0.5 mL of argon and 0.1 μL of krypton (39). This liter of "modern" water will therefore contain $\sim 1 \times 10^4$ [39]Ar atoms and $\sim 1 \times 10^3$ [81]Kr atoms. Taking their respective half-lives into account this corresponds to a decay rate of $3 \times 10^{-3}$ hr$^{-1}$ for [39]Ar and $4 \times 10^{-7}$ hr$^{-1}$ for [81]Kr. These extremely low rates have pushed not only the analytical techniques, but also the sampling techniques, to their technical limits. For [85]Kr the situation is quite different. One liter of "modern" ocean water contains $\sim 2 \times 10^4$ [85]Kr atoms, corresponding to a decay rate of 0.1 hr$^{-1}$. Decay counting of [85]Kr is performed at several laboratories, and it is a challenge to any alternative methods to further reduce both the sample size and measurement time for routine environmental applications.

In the following sections we discuss three promising analytical methods: Low-level Counting (LLC), Accelerator Mass Spectrometry (AMS) and Atom Trap Trace Analysis (ATTA).

### 3.1. Low Level Counting (LLC)

For decades the application of noble gas radionuclides was based on their radioactivity measurement by detecting emitted β-particles and x-rays. This method is preferred if the half-life is relatively short and, consequently a considerable fraction of the atoms present in a sample decays during the measurement. For example, in one week of counting, approximately $1 \times 10^{-3}$ of the [85]Kr atoms in the sample decay, the decay fraction is $5 \times 10^{-5}$ for [39]Ar and only $6 \times 10^{-8}$ for [81]Kr. LLC measurements are preferably performed in a low-background environment. One such state-of-the-art laboratory is located 35 meters below the surface in an underground laboratory at the University of Bern in Switzerland. This laboratory has pioneered LLC measurement on noble gas radionuclides such as [39]Ar (40) and [81]Kr (9), and has since performed the bulk of LLC measurements on $^{37}$Ar, $^{39}$Ar, and $^{85}$Kr. The following discussion on LLC is therefore primarily based on the experience of this laboratory.

Most of the measurements performed with noble gas radionuclides involve the extraction of dissolved gases from water. Since large volumes of water have to be sampled, degassing of the water samples is done in the field. The fact that isotope ratios are determined ($^{39}$Ar/Ar and $^{85}$Kr/Kr) has the big advantage that an incomplete degassing yield can be accepted. For $^{39}$Ar measurements, the sample size of 1500-5000 liters water is driven by the amount of argon required (0.5-2 L), the argon content in water (0.4-0.5 mL/L) and the degassing efficiency (70-80%). The volume of extracted gas is very variable because of variable amounts of $O_2$, $CO_2$ and $CH_4$ dissolved in the water. The composition of the extracted gases has to be measured before the separation of Kr and Ar because different procedures are necessary for different gas compositions. For gas proportional counting (see below) high purity argon and krypton is needed. In the argon fraction the krypton content has to be reduced by a factor of at least $10^4$ to eliminate the background induced by present-day atmospheric $^{85}$Kr activity. Extracted gas samples with a total volume between 10 liters and several hundred liters are processed in a multi-step purification system.

$^{39}$Ar decays by β-radiation with a maximum energy of 0.56 MeV without γ-emission. Therefore gas proportional counting (Fig. 3) is appropriate for the identification of $^{39}$Ar decay. This argument also holds for $^{85}$Kr (maximum β-energy = 0.69 MeV), because at the concentrations present in environmental samples the probability to detect the small γ-radiation branch (0.4% per decay of $^{85}$Kr) is too low. Because of the relatively high β-energy of these two isotopes, the whole energy spectra above certain electronic cut-off can be used (Fig. 4). (The identification of $^{37}$Ar activity is also possible by gas proportional counting).

The deposition of the β-energy within the gas volume is higher if a high gas pressure is used. In addition, a smaller volume of the counters reduces the background because the surface and therefore the influence of self activity of the counter material (mostly high purity copper) are smaller. Gas proportional counters of different volumes (10-100 mL), pressures (10-23 bar), and methane admixtures (5-10%) are used at Bern (41 – 44). Depending on the configuration, counting times for $^{39}$Ar range from 8 days to 60 days, and for $^{85}$Kr from 3 days to 6 days.

LLC was also used for the first detection of $^{81}$Kr in the atmosphere (9). A one-liter counter was filled with 2 bars of pre-bomb krypton, i.e. without $^{85}$Kr activity; and 15% $CH_4$ was added. The identification of the nuclide was possible by observing the characteristic 13.5 keV x-ray peak, induced by the electron-capture decay. However, due to large content of anthropogenic

$^{85}$Kr in the atmosphere (Fig. 2), even a very small admixture of modern krypton makes it impossible to count the decay of natural $^{81}$Kr nowadays.

The lowest measured $^{85}$Kr activities in groundwater samples are on the order of 0.2 dpm/mL Kr. Such values and even somewhat larger ones do not necessarily indicate the presence of an admixture of recent water or air. A small contamination during sampling and sample processing, corresponding to about 100-mL of modern air within e.g. a sample of 200 liters of extracted gas, is quite possible in practice. $^{85}$Kr is therefore more sensitive for the indication of a contamination than $^{39}$Ar. For the detection of a small amount of admixed recent water, the $^{85}$Kr activity measurement should be combined with $^3$H, $SF_6$, $^3$He/$^3$H, CFCs results and the $O_2$-content in the extracted gases.

## 3.2. Accelerator Mass Spectrometry (AMS)

When analyzing long-lived isotopes, atom counting has a number of advantages over decay counting. The efficiency and speed of atom counting is not fundamentally limited by the long half-lives of isotopes, nor is it affected by radioactive backgrounds in the environment or in samples. LLC was used for $^{14}$C-dating for several decades until it was replaced by a more efficient atom-counting method: accelerator mass spectrometry (45). AMS quickly developed into a versatile tool to measure long-lived cosmogenic radionuclides in many different archives (4, 46-51).

In an AMS analysis, ions are first produced from a sample in an ion source, and are then passed through a first set of low-energy analyzing magnets in order to select those with the correct charge-to-mass ratio. The selected ions are accelerated and, in some cases, passed through a stripper foil which largely removes molecular isobars. The surviving ions are passed through an additional series of magnetic and electrostatic analyzers before being further analyzed and counted in a detector system. At high energy (MeV–GeV), more versatile and discriminatory ion detection techniques such as energy-loss and time-of-flight measurements can be applied to help identify the nuclei of the ions and further reduce the influence of isobaric interferences. The high sensitivity and selectivity of the method makes it possible to measure a trace nuclide at a rate of several counts per hour from a μA primary beam (~$10^{16}$ ions per hour).

A typical AMS apparatus uses a tandem accelerator that starts with a negative ion beam (47). Several AMS laboratories also employ linacs and cyclotrons, which have the added advantages that the accelerator provides additional mass selection for a given charge state and that they can work with isotopes that do not form negative ions, as in the case of noble gas isotopes.

### 3.2.1. $^{81}$Kr, full stripping

The main technical problem for the detection of $^{81}$Kr with AMS resides in the separation from its isobaric, stable and abundant $^{81}$Br background. This was recently overcome by Collon et al. (21) using the full stripping technique (Fig. 5). Once all of the atomic electrons are removed, $^{81}$Kr$^{+36}$ (Z=36) can be cleanly separated from $^{81}$Br$^{+35}$ (Z=35). The AMS measurements of $^{81}$Kr in atmospheric krypton were performed using the K1200 Superconducting Cyclotron of Michigan State University, which accelerated ions to an energy (3.6 GeV) high enough for efficient full stripping. After the acceleration in the cyclotron, the $^{81}$Kr$^{17+}$ions were passed through a beryllium foil (18.8 mg/cm$^2$), which provided a 70% stripping efficiency to reach the 36+ charge state. The fully-stripped $^{81}$Kr$^{36+}$ were then separated from the $^{81}$Br$^{35+}$ using the A1200 mass spectrometer, and detected with two $\Delta$E-E telescope detectors. Additional time-of-flight information was provided by a scintillator placed after the first bending magnet of the spectrometer. A typical time-of-flight vs. $\Delta$E spectrum obtained using this setup is shown in Fig. 6. For normalization, the source output current ($^{84}$Kr$^{17+}$and $^{86}$Kr$^{17+}$) was measured in a Faraday cup placed after the ion-source analyzing magnet.

The nuclear reaction products induced by the high $^{81}$Br$^{17+}$ background in the stripper foil heavily dominated the final spectrum, and had to be suppressed by several orders of magnitude before $^{81}$Kr in natural samples could be detected. As krypton and bromine have very different melting temperature of -156.6°C and -7.2°C, respectively, bromine in the sample was suppressed by freezing the sample with liquid nitrogen. In addition, tests revealed that the largest suppressor of $^{81}$Br was the support gas used in the ion source. By using argon as support gas, reductions in the bromine contamination by two orders of magnitude were obtained (21). These steps opened the door to possible measurements of $^{81}$Kr below the natural level of $5\times10^{-13}$. This AMS setup was used to perform $^{81}$Kr-dating of groundwater samples (52) and to compare pre-bomb and post-bomb abundances of atmospheric $^{81}$Kr/Kr (24).

One very important condition for AMS measurements of noble gases is the availability of two separate ECR sources at the injection end of the accelerator. The two-source arrangement greatly reduces cross contamination between the enriched samples with elevated isotopic ratios, necessary for tuning and setting up of the accelerator and detectors, and the natural samples in subsequent measurements.

### 3.2.2. $^{39}$Ar, gas-filled spectrograph

The full stripping technique works well only when the trace isotope has a higher atomic number Z than that of contaminant isobars. This, however, is not the case for the $^{39}$Ar (Z=18) – $^{39}$K (Z=19) isobar pair, or the $^{85}$Kr (Z=36) – $^{85}$Rb (Z=37) pair. In the case of $^{39}$Ar analyses using AMS, the isobaric separation was achieved using the gas-filled spectrograph technique (53) (Fig. 7). The two key effects combined in this technique, the bending of the ion's trajectory in the magnetic field and the loss of ion energy in the gas, have different relations to the charge and mass of the ion, thus allowing the separation of both isotopes and isobars. In particular, in the gas-filled region of the magnet, the discrete trajectories of each of the charge states of the ions coalesce around a trajectory defined by the mean charge state of the ion in the gas (54).

The development of this method and the $^{39}$Ar/Ar measurements that followed were performed at Argonne National Laboratory using the ATLAS accelerator facility. $^{39}$Ar$^{8+}$ ions were alternately produced in ECR I and ECR II sources, accelerated to 232 MeV by the linac, separated from their isobar $^{39}$K$^{8+}$ ions in the gas-filled spectrograph, and finally detected using a position sensitive parallel plate avalanche counter (PPAC) followed by an ionization counter (55, 56) (Fig. 7). This detector has a vertical opening of 100 mm, which matches the width of the beam spot on the focal plane of the gas-filled magnet and has an improved acceptance by a factor of 2.5, compared to the standard focal plane detector. Although a clear separation was obtained between the $^{39}$K and $^{39}$Ar peaks, the intensity of the $^{39}$K rate created signal pile-up in the detector (Fig. 8). This pile-up became an even bigger problem when trying to measure samples with below atmospheric $^{39}$Ar concentrations. This problem also limited the $^{40}$Ar$^{8+}$ beam current and the final $^{39}$Ar count rate as the $^{39}$K rate scaled with the $^{40}$Ar$^{8+}$ beam current. It was therefore imperative to further reduce the $^{39}$K background by at least one order of magnitude. One of the major sources of $^{39}$K in the source was suspected to be the plasma chamber liner. In order to obtain a "clean" plasma chamber a quartz liner was installed in the ECR ion source. This improvement had the effect of decreasing the $^{39}$K background by a factor of 130 (57) without substantially affecting the intensity of the $^{40}$Ar$^{8+}$ beam (78 eµA), and made it possible to measure $^{39}$Ar/Ar at and below the natural level. The detection limit for $^{39}$Ar/$^{40}$Ar, determined from measurements performed on a half-million-year old water sample, is at $4\times10^{-17}$, corresponding to the analysis limit on age at 1200 years. The overall counting efficiency has reached $3\times10^{-3}$ (57), and was calibrated by analyzing atmospheric argon samples with known $^{39}$Ar/$^{40}$Ar ratio of $8.1\times10^{-16}$, as well as two neutron-activated samples with higher $^{39}$Ar/$^{40}$Ar values.

The $^{85}$Kr–$^{85}$Rb case is similar to the $^{39}$Ar–$^{39}$K case. An initial exploratory experiment using the identical experimental setup has started at Argonne National Laboratory. It should be noted that the gas-filled spectrograph technique would not work for the case of $^{81}$Kr. The $^{81}$Kr counts would end up on the low-energy side of [81]Br counts, and unavoidable tailing to lower energy and lower magnetic rigidity from [81]Br would mask the real [81]Kr counts.

### 3.3. Laser-based methods

During the past three decades a number of methods based upon laser spectroscopic techniques were proposed and developed (58). The high degree of selectivity obtained by these methods is a result of resonant laser-atom interaction. Atoms of different elements interact resonantly with light at significantly different frequencies due to difference in their atomic structure. Atoms of different isotopes of the same element exhibit isotope shift, a small change in their resonance frequencies caused by the variation in the nuclear mass and charge radius. By tuning the laser frequency precisely onto the resonance of a particular isotope, one can selectively excite, ionize, or manipulate the atoms of this isotope while having a much smaller effect on the other isotopes and elements.

The selectivity ($s$) of a single optical excitation is defined here as the ratio of the probability of exciting the selected isotope over the probability of exciting a neighboring (contaminant) isotope and, in a simple model that assumes Lorentzian lineshape, can be expressed as

$$s \approx 4 \times \left( \frac{\Delta}{\Gamma} \right)^2 \quad \text{when } \Delta >> \Gamma,$$

where $\Delta$ is the frequency difference between the isotopes, and $\Gamma$ is the interaction linewidth. While a single resonant excitation can possess an adequate elemental selectivity due to the usually large frequency differences ($\Delta/\Gamma \sim 10^8$) between different elements, its isotopic selectivity ($\Delta/\Gamma \sim 10^2$) is not sufficient. In order to overcome this deficiency, two laser-based methods with additional features were proposed in the late 1970's:

*Resonance ionization mass spectrometry (RIMS)* (59, 60). In RIMS, a high overall selectivity is achieved by combining the elemental (isobaric) selectivity of laser excitation and the isotopic (mass) selectivity of mass spectrometry. Here an atom is selectively excited and ionized via resonant laser light, the ion is then passed through a mass filter and detected by an ion counter such as a channeltron. In theory, an isotopic selectivity of $s_{laser} \sim 10^4$ can be reached at the resonance excitation stage, and $s_{mass} \sim 10^8$ can be reached at the mass spectrometry stage, resulting in an overall isotopic selectivity of $S = s_{laser} \times s_{mass} \sim 10^{12}$. In practice, however, any line broadening and non-resonant excitation effects due to atom-atom or atom-electron collisions limit the selectivity at a lower level. RIMS was used to count [81]Kr atoms in atmospheric and groundwater samples with an efficiency of over 50%, but only after the samples had gone through several stages of isotope enrichment (61 – 63). RIMS was used in early attempts of dating groundwater and ice (64, 65). After two decades of effort, the whole procedure proved to be too complicated to perform practical and reliable analyses. On the other hand, RIMS has been successfully applied to analyze many trace-isotopes with higher abundance and has been firmly established as a general analytical method.

*Photon-burst mass spectrometry (PBMS)* (66) – Detecting a single atom by observing its fluorescence induced with resonant laser excitation is in general difficult because of the low efficiency in photon collection (~ 10%) and photon counting (~ 10%) as well as the high background due to either scattered light or dark counts of the detector. These difficulties can be overcome by inducing from each selected atom not only one, but $10^2$ or even more photons in a short burst. The selectivity increases exponentially with the number of photons detected in a single burst as the overall selectivity (*S*) of a multi-step excitation process is simply the product of the selectivity (*s*) of each step, $S = s^n$, here *n* is the number of excitations. As in the case of RIMS, selectivity can be further improved by combining the photon-burst detection with mass spectrometry to form PBMS, which was used to detect $^{85}$Kr at the isotopic abundance level of $10^{-9}$ (67), about two orders of magnitude above the atmospheric level. At present, the primary limiting factor is the difficulty of producing and detecting a large enough photon burst during the short interaction time (~ μs) as the atoms move across the laser beam. While in principle this problem can be solved by lengthening the interaction region and adding more photon detectors, in practice such a solution has been difficult to implement and is still under development. In the following section we discuss a recently developed method that dramatically increases the interaction time by probing atoms confined in a trap.

### 3.3.1. Atom Trap Trace Analysis (ATTA)

In ATTA, an atom of a particular isotope is selectively captured by a laser trap and then detected by observing its fluorescence while it is in the trap (68). The general techniques of laser cooling and trapping have already been well reviewed in other articles and books (69, 70). Here we focus on the specifics of these techniques in the context of using them for trace analysis applications.

A magneto-optical trap (MOT) (71), a type of robust and efficient atom trap, is used here as a selective concentrator. When the laser frequency is tuned to within a few natural linewidth on the low-frequency side of the resonance of the targeted isotope, only atoms of this particular isotope are trapped. Atoms of other isotopes are either deflected before reaching the trap or are allowed to pass through the trap without being captured. The detection method here takes full advantage of the high selectivity of photon burst spectroscopy. An atom can be trapped and observed for 100 ms or longer, during which $10^6$ fluorescence photons can be induced from a single trapped atom and as many as $10^4$ photons can be detected. Furthermore, the viewing region where the atoms are trapped and from which the fluorescence is collected is small (< 1 mm across), therefore, spatial filtering can be implemented to reduce the background photons scattered off the windows and walls. These advantages allow the counting of single atoms to be done with a high signal-to-noise ratio and, since the selectivity depends exponentially on the number of photons detected in a burst, also ensure a superb selectivity. Indeed ATTA is immune to contamination from other isotopes, elements, or molecules. The selectivity of ATTA is only statistically limited by the atomic beam flux, more specifically, by the number of atoms that can be sent through the trap in a practical time.

MOT's have so far been applied to approximately 20 elements. Atoms that can be confined by a MOT must have a cycling or a quasi-cycling transition. This effect preferably selects atoms with few uncoupled electrons and simple ground-level structure, such as alkali, alkaline earth and noble gas elements.

### 3.3.2. $^{81}$Kr and $^{85}$Kr, ATTA

An ATTA system developed at Argonne National Laboratory has been used to analyze $^{81}$Kr and $^{85}$Kr in natural samples (68, 72). With a modern atmospheric Kr sample, the typical count rate is 12 hr$^{-1}$ for $^{81}$Kr and 240 hr$^{-1}$ for $^{85}$Kr, the counting efficiency is $1\times10^{-4}$, and the required sample size is 50 µL of Kr. This system has met the requirements of implementing practical $^{81}$Kr-dating of old groundwater. On the other hand, the atomic beam flux achieved in this system is still too low for the analysis of the much less abundant $^{39}$Ar.

Fig. 9 shows the schematic of the system. Either a Ti:Sapphire ring laser system or a diode laser system can be used to generate the 811 nm light needed to trap Kr atoms in the metastable $5s[^{3}/_{2}]_2$ level by repeatedly exciting the $5s[^{3}/_{2}]_2$ - $5p[^{5}/_{2}]_3$ cycling transition. The vacuum system consists of the following components: the sample reservoir, the source chamber, the transverse cooling chamber, and the trap chamber. The system is differentially pumped by three turbo-pumps so as to maintain a pressure of ~ $10^{-3}$ Torr in the source chamber and a pressure of ~ $10^{-8}$ Torr in the trap chamber. A getter pump in the source chamber removes the reactive gases, such as hydrogen, water, etc., from the vacuum system while leaving the krypton sample intact. A novel feature of the system is that it can be switched into a mode that recirculates the krypton atoms repeatedly through the system and dramatically improves the counting efficiency.

In an analysis, a krypton gas sample is injected into the system through a discharge region, where approximately $1\times10^{-3}$ of the atoms are excited into the metastable level $5s[^3/_2]_2$ via collisions with the energetic electrons and ions. The thermal (300 K) atoms are then transversely cooled, decelerated with the Zeeman slowing technique, and captured into a MOT. The MOT can capture the abundant $^{83}$Kr (isotopic abundance = 11.5%) at the rate of $1\times10^9$ s$^{-1}$. Since an atom typically lasts 0.5 s in the trap before being bumped out of the trap in collisions with thermal background atoms or molecules, the rare atoms are not accumulated in the trap; instead they are usually captured and counted one at a time. A single trapped atom can be unambiguously observed with a signal-to-noise ratio of 40 by detecting its strong fluorescence induced by the trapping laser beams (Fig. 10).

The capture rates of $^{81}$Kr and that of a stable krypton isotope $^{83}$Kr differ by 11 orders of magnitude. At present, the $^{81}$Kr/$^{83}$Kr ratio cannot be measured directly due to inadequate dynamic range in the ATTA detector system. Instead, a controlled amount of $^{85}$Kr is introduced into the sample, and the $^{81}$Kr/Kr ratio is determined based upon two measurements: the $^{85}$Kr/$^{81}$Kr ratio (in the range of 1-100) measured using ATTA and the $^{85}$Kr/Kr ratio measured using LLC. In order to demonstrate the validity of ATTA as a quantitative analytical method and to calibrate the counting efficiency, measurements were performed on ten young (age < 100 yr) atmospheric samples, in which the $^{81}$Kr/Kr ratios are expected to be identical (72). The results showed that the $^{85}$Kr/$^{81}$Kr ratios measured using ATTA and the $^{85}$Kr/Kr ratios measured using LLC were indeed proportional to each other within ±10%, and that the $^{81}$Kr/Kr ratios of pre-bomb and post-bomb samples are identical within ±8%.

### 3.4. Comparison of LLC, AMS, and ATTA

In table 1, we compare the conditions achievable for the detection of noble gas radionuclides with the three different methods discussed above. Compared to decay counting, atom counting is clearly superior for $^{81}$Kr analyses, but it is also very competitive for $^{39}$Ar and $^{85}$Kr analyses where the smaller sample size needed for AMS and ATTA presents a significant advantage in measurements on environmental samples.

## 4. APPLICATIONS

### 4.1. Groundwater

Groundwater is an increasingly important resource for our industrial, agricultural, and household needs. Worldwide studies of groundwater have been carried out in the past few decades with the aim of achieving sustainable use and preserving the quality of groundwater resources. Dating, or determining the residence time of groundwater, with environmental tracers has been used extensively to help uncover the history of recharge, i.e. to find out at which location the water was injected into the underground aquifer and when the recharge happened, and to map the direction and speed of flow over the aquifer region (73). Such information is essential in managing the use of groundwater as well as in understanding the subsurface transport of both chemical and nuclear pollutants. Moreover, paleoclimate information can be extracted from the analysis of groundwater with known residence time (74, 75). Here residence time is defined as the time since the water sample was isolated from the atmosphere. This concept can be complicated by the fact that water of different residence times can mix, in which case several tracers sensitive to different time scales should be used to untangle the multi-component history of the sample.

Due to their uniform and stable atmospheric distribution and simple chemical properties, noble gas radionuclides offer great advantages over the other environmental tracers. $^{85}$Kr is ideal for dating groundwater less than 50 years old, a period most relevant to assessing the subsurface transport of chemical pollutants. It competes against other man-made tracers such as $^3$H-$^3$He, CFC and SF$_6$, all of which can be analyzed with conventional mass spectrometry with a small sample of less than 0.1 L of water (76). However, the atmospheric input of $^3$H depends strongly on season and latitude, and continues to decline over time; molecular tracers tend to suffer from the uncertainties due to subsurface chemical degradation. Beyond the reach of these young water tracers, the age range of 50–1000 years can be addressed with $^{39}$Ar dating and $^{14}$C dating. The two methods compliment each other: $^{14}$C dating can be complicated by subsurface carbonate chemistry, and $^{39}$Ar dating by the subsurface production via the $^{39}$K(n, p)$^{39}$Ar reaction. Even for dating water samples much older than 1000 years, both $^{85}$Kr and $^{39}$Ar dating are useful in investigating the mixing-in of younger water. For water older than 50,000 years, beyond the reach of $^{14}$C dating, $^{81}$Kr is the ideal tracer. Although much work has been done with another long-lived isotope, $^{36}$Cl, a tracer that can be measured routinely with AMS, the $^{36}$Cl/Cl results are hindered by significant uncertainties in both the $^{36}$Cl input function and subsurface chlorine chemistry (17).

### 4.1.1. $^{81}$Kr dating of the Great Artesian Basin in Australia with AMS

The development of AMS analyses of $^{81}$Kr culminated in the first demonstration of $^{81}$Kr dating of old groundwater. The sampling site, situated in the south-west segment of the Great Artesian Basin in Australia, was selected for three reasons: (a) based on previous hydrodynamic studies as well as $^{36}$Cl data the existence of very old groundwaters in an age range accessible by the [81]Kr method was proposed (77); (b) the Artesian condition has the advantage that water samples reach the surface under their own pressure without pumping, which minimizes the danger of contamination with atmospheric air during the sampling process; (c) the Great Artesian Basin is the largest artesian groundwater basin in the world with water flowing in sandstones covered by extensive layers of shale. It provides water not only for the local population and agriculture in the arid interior of the Australian continent but also for major industries (copper and uranium mining) with rather large demand for processing water.

Using the cyclotron technique discussed in Section 3.2.1, Collon *et al.* (52) determined the residence time of groundwater extracted from four wells in the Great Artesian Basin to be $(4.0\pm0.5)\times10^5$ yr, $(3.5\pm0.5)\times10^5$ yr, $(2.9\pm0.4)\times10^5$ yr and $(2.3\pm0.4)\times10^5$ yr. In a typical measurement, about 350 liters of gas were stripped from 16 tons of water in the field next to the wells. The gas was compressed into suitable pressure bottles of 20 L volume, and was then taken to the laboratory to be purified to obtain 0.5 mL of krypton. This sample material containing about $2\times10^6$ [81]Kr atoms was injected into the ion source of the cyclotron and resulted in about 60 [81]Kr counts in 9 hours of measurement time (Fig. 6). The gas processing efficiency was approximately 50%, and the AMS counting efficiency was $3\times10^{-5}$. This demonstration is significant as it was the first time that old groundwater was definitively dated. These [81]Kr ages were then used to calibrate the [36]Cl dating method (18).

### 4.1.2. [81]Kr dating of the Nubian Aquifer in Egypt with ATTA

More recently ATTA was applied to [81]Kr dating of old groundwater (78). Measurements of [81]Kr/Kr in deep groundwater from the Nubian Aquifer in the Western Desert of Egypt were performed (Fig. 11). The measured [81]Kr ages range from $\sim2\times10^5$ to $\sim1\times10^6$ yr, correlate with [36]Cl/Cl ratios, and are consistent with lateral flow of groundwater from a recharge area near the Uweinat Uplift in southwest Egypt. Moreover, low $\delta^2H$ values of the [81]Kr-dated groundwater reveal a recurrent Atlantic moisture source during Pleistocene pluvial periods.

Samples of Nubian Aquifer groundwater extracted from wells were measured for [81]Kr, [36]Cl, and other chemical and isotopic constituents. For [81]Kr dating, dissolved gas was extracted from several tons of water in the field at six sites. The [81]Kr data indicate that ages increase progressively along flow vectors predicted by numerical hydrodynamic models, verifying distant lateral flow of deep groundwater toward the northeast from a recharge area southwest of Dakhla. Furthermore, the [81]Kr data indicate relatively high flow velocities (~2 m/yr) from Dakhla toward Farafra, and low velocities (~0.2 m/yr) from Dakhla toward Kharga and from Farafra to Bahariya. These observations are consistent with the areal distribution of hydraulically conductive sandstone within the aquifer and they provide support to some of the existing hydrodynamic models. Southwestward extrapolation of the ~2 m/yr flow rate inferred from the difference in $^{81}$Kr ages for Dakhla and Farafra is consistent with recharge in the area of the Uweinat Uplift near the Egypt-Sudan border (Fig. 11). In this area, the Nubian sandstone is exposed (or buried beneath sand sheets or dunes) at elevations between 200 and 600 m above sea level over a wide area, forming a broad catchment for recharge of the Nubian Aquifer.

## 4.2. Ocean water

It is now apparent that the oceans play a major role in the Earth's climate system by transporting the enormous amount of heat contained in the water around the globe. One of these pathways is the Atlantic Conveyor Belt system where warm surface water from equatorial regions cools in the Far North Atlantic and sinks to greater depths. In a simplified picture one can say that the cold deep water flows southward in the Atlantic, around the southern end of Africa, across the Indian Ocean into the Pacific Ocean where it rises to the surface and flows back to the North Atlantic (Fig. 12). The whole cycle takes about 2000 years. The amount of heat transported from the Equator to the polar regions by the oceans is comparable to the amount transported by the atmosphere. Moreover, because the ocean has such an enormous heat capacity, it moderates the climatic changes occurring in the atmosphere, for example by trapping additional heat generated in the greenhouse gases (79). With its half-life of 269 years and its favorable geochemical properties, $^{39}$Ar is an ideal radionuclide to study this important circulation system in oceanography (33, 80). $^{39}$Ar would close the gap between the time scales covered by transient tracers such as $^{3}$H ($t_{1/2}$ = 12.3 y), CFCs (quasi-stable) and $^{14}$C ($t_{1/2}$ = 5730 y). Moreover, noble gas exchange between the surface ocean water and the atmosphere happens much faster than the exchange of $CO_2$. Surface ocean equilibrium time for $^{39}$Ar is approximately 30 days, whereas for $^{14}$C it takes a few years (80). It should also be noted that in the 20–1000 year range considered for applications in ocean circulation studies, the $^{39}$Ar concentration varies between 100% and 10% of its modern level, whereas $^{14}$C varies only from 100% to about 84%. Furthermore, by combining different isotopes with different equilibration times and half-lives, a great deal of additional information can be obtained on the "history" of a volume of ocean water including the mixing and diffusion processes (80 – 82). One of the recent puzzles in the flow of ocean water is the possible slowdown of the Southern Ocean Deep Water formation (83), which may significantly affect temperature changes in our time. The measurement of robust natural tracers such as $^{39}$Ar may well contribute decisively to our understanding of these complex processes.

### 4.2.1 [39]Ar dating of Atlantic Ocean with LLC and AMS

Between 1979 and 1991, ocean water samples of ~1000 liter size were taken for [39]Ar measurements during several international research cruises. LLC measurements of the samples at the University of Bern have resulted in the first series of [39]Ar data sets in oceanography (84). The [39]Ar data provided information on the aging, circulation and mixing processes, primarily in the deep ocean (85). The measured [39]Ar activity values cover from 100% to 10% modern, which is particularly important in deep southern oceans where the $\delta^{14}C$ variations are small. The [39]Ar data indicate that the North Atlantic Deep Water needs an average of 140±40 years to flow from its origin southwards to the Equator; this result agrees with the result reached independently using [14]C dating. Furthermore, the amount of water transported by this current as well as the flow speed can be calculated based on the [39]Ar data.

AMS measurements of the [39]Ar concentrations in several argon samples extracted from ocean water as well as groundwater sample have been successfully performed at Argonne National Laboratory. With an overall efficiency of ~ $3\times10^{-3}$, AMS measurements of [39]Ar/[40]Ar isotope ratios in the range of $10^{-16}$ were performed (57) on three ocean samples collected during the South Atlantic Ventilation Experiment (SAVE). One of the samples off the eastern coast of Brazil was from 17 degrees south of the equator and from a water depth of 4700 m (Fig.12). It gave a [39]Ar/[40]Ar ratio that was 32 percent of the modern atmospheric ratio. Two samples off the coast of Argentina at 47 degrees south gave ratios of 65 and 44 percent modern argon at water depth of 850 and 5000 m, respectively. The precision of these measurements was determined by counting statistics and ranged from 15% to 25% (larger errors for the "older" samples). While improvements on accelerator transmission and beam current (and therefore [39]K background reduction) are still needed before any serious attempts can be made for measuring ocean water sample at precisions below 5% (in particular for "old" samples), we believe that the reported results represent a first indication that AMS analyses of [39]Ar/[40]Ar ratios with precisions below 5% in less than 10 hours of measuring time, and with ocean water samples of 20 L size, may be achievable.

### 4.2.2 [85]Kr in oceanography

The LLC analysis of [85]Kr/Kr in ocean water requires a 20-$\mu$L sample of krypton that has to be extracted from about 400 L of water (86). Although it is possible to collect 400 L of water from the ocean and from high yielding groundwater wells, processing and analyzing the samples is still labor intensive, requiring about two man-days per sample. In practice, the equipment needed to collect 400-L seawater samples is not available on most oceanography research ships.

These difficulties have so far prevented the wide use of $^{85}$Kr in oceanography, but the measurements that have been made clearly show its potential. Chlorofluorocarbons (CFCs), anthropogenic gases that have entered the natural water systems from the atmosphere and have increased dramatically in the atmosphere since the 1950s, are much easier to measure and are indeed widely applied in oceanography. However, the interpretation of CFC data often presents problems because CFCs decompose when the water becomes anoxic. This would not be a problem for $^{85}$Kr.

There are also special problems in oceanography that can be uniquely addressed with $^{85}$Kr. One such problem is detecting leakage from nuclear wastes dumped in the ocean. The Former Soviet Union dumped a large quantity of liquid and solid nuclear wastes, including fueled reactors, in the Barents and Kara Seas of the Arctic Ocean (87). Concerns about the possibility of radioactive materials leaking from these dump sites and being transported to other parts of the Arctic Ocean, including the northern slope of Alaska, has prompted investigations by the Arctic Nuclear Wastes Assessment Program (88). $^{85}$Kr would be one of the first substances to leak from solid wastes such as fuel rods or fueled reactors and could provide an advanced warning that other materials would soon leak. Approximately 40 $^{85}$Kr measurements taken in the Arctic Ocean suggest that such leakage may indeed be occurring (89) and no significant leakage of other radionuclides from these dump sites has so far been detected (87).

### 4.3 Ice cores

The large ice sheets of Greenland and Antarctica are the main store-houses of ice on earth. Approximately 2.1 % of all water on Earth is stored there (90), with more than double that amount land-locked during the Last Glacial Maximum some 20,000 years ago. Large ice-core drilling projects have been pursued during the last 30 years in Greenland and Antarctica, retrieving ice records back to 250,000 years (91, 92) and 420,000 years (93), respectively. New ice core drilling projects such as the European Project for Ice Coring in Antarctica (EPICA) may reach back in time beyond 500,000 years.

Ice cores are the most important archives to study the composition of the atmosphere back in time. Through precipitation and air bubble occlusion a direct imprint of atmospheric conditions is preserved in time. Analysis of air bubbles revealed that the $CO_2$ content never exceeded 300 ppm in the past 420,000 years (93), except for the last 150 years where it rapidly increased to a current value of 370 ppm (94). In the same time period the $CH_4$ concentration has increased from 700 to 1700 ppb (95). A wealth of other information can be extracted from ice cores, such as the paleotemperatures through $\delta^{18}$O measurements (96), dust concentrations indicating drier periods, acidity peaks linked to major volcanic eruptions, etc. While the polar ice sheets record climatic conditions at high latitudes, in recent years lower-latitude ice cores are also being explored from glaciers of high-altitude mountains such as Kilimanjaro (97).

Ice cores can be dated most accurately as long as annual layers can be counted back in time (similar to counting tree rings). This can be accomplished by various methods such as annual grey-scale variations and seasonal $\delta^{18}O$ oscillations. However, eventually the annual structure disappears under the enormous pressure of the overlaying ice, and ice accumulation models have to be employed to determine the age of deep ice. In principle, $^{81}Kr$ with a half-life of 230,000 years would be well suited to date ice back to one million years, and perhaps beyond. The main obstacle so far is the low $^{81}Kr$ concentration in ice (~1000 $^{81}Kr$ atoms per kilogram of modern ice) combined with only small amounts of ice (a few kg) available from deep ice cores. The counting of such small amounts of $^{81}Kr$ atoms requires an extremely high efficiency (at least 10%), currently beyond the capabilities of both AMS and ATTA.

## 5. OUTLOOK

In conclusion, significant progress has been made in the development of practical methods of analyzing the three important noble gas radionuclides: $^{39}Ar$, $^{81}Kr$ and $^{85}Kr$. As is summarized in table 1, AMS can now be used to analyze $^{39}Ar$ in a few tens of liters of water depending on ages, and ATTA can be used to analyze $^{81}Kr$ in a few tons of water and $^{85}Kr$ in approximately 100 liters of water. With these advancements, the geological proof-of-principle applications of these isotopes have been realized, thus opening ways towards more geological applications in the near future based on the presently available methods.

Among the leading methods, LLC is mature and is unlikely to see further significant advances. In contrast, both AMS and ATTA are newly developed methods and, in our belief, have much room for improvement in both counting speed and sample size, as well as in ease of use.

For AMS the major challenge is the suppression of isobaric background. Currently, big accelerator facilities such as the Superconducting Cyclotron at Michigan State University ($^{81}Kr$) and ATLAS at Argonne National Lab ($^{39}Ar$) are required to reach the high energy needed for isobar separation. Desirable overall efficiencies (at least for $^{39}Ar$ analysis in ocean water samples) are within reach, perhaps requiring a dedicated ECR source to improve things further. However, it is difficult to imagine using these large facilities primarily devoted to nuclear physics for "routine" measurements of large sample numbers, as is quite common for $^{14}C$ measurements at small AMS facilities around the world. It is therefore important to find scientific questions in Earth sciences where these more elusive radionuclides can make an impact with a limited number of measurements.

For ATTA, we estimate that an improvement by a factor of 10 in both counting speed and efficiency is possible by further optimizing the system and with more laser power. For more dramatic improvements, the present practice of producing metastable krypton atoms (Kr*) with a discharge, which induces a loss factor of $10^4$ in both counting speed and single-pass efficiency (without recirculation), must be replaced. An alternative method of using pure optical excitation schemes to produce Kr* has been realized (98). The flux and efficiency of such an optical source is currently under investigation.

If and when an atom-counting method such as AMS or ATTA is improved to the point that analyses of these noble gas radionuclides can be carried out quickly and routinely with water samples of 100 L or less so that geologists can sample water directly rather than performing the lengthy and somewhat tricky gas-stripping procedure in the field, we expect to see a dramatic increase in the use of these isotopes. The counting efficiency required to analyze 100 L of water with a statistical precision of 5% is approximately 0.1% for both $^{85}$Kr and $^{39}$Ar, and 1% for $^{81}$Kr. The ultimate challenge is to perform $^{81}$Kr dating of deep ice core samples, which may be available at the size of 1 L or less. Its realization would require a method with a counting efficiency of at least 10%.

Of course it would not be wise to rule out the possibility of the emergence of a brand new method that tops all the existing methods. It is instructive to look back at the history of the present methods. Since Libby and co-workers demonstrated $^{14}$C dating with LLC in 1950, counting $^{14}$C atoms with conventional mass spectrometry was pursued for more than two decades, with its selectivity improved slowly and steadily towards the natural level, until 1977 when AMS emerged suddenly and succeeded in the dream of counting $^{14}$C atoms at the natural level. It is interesting to note that laser methods followed a similar pattern of development. Counting $^{81}$Kr and $^{85}$Kr atoms with two early laser methods, RIMS and PBMS, were pursued for over two decades, with much progress and some limited successes, until 1999 when counting both krypton isotopes at the natural levels was first demonstrated with the use of a new method ATTA.

**ACKNOWLEDGEMENT**

We would like to thank H. H. Loosli and R. Purtschert for providing key information on the chapters of Low Level Counting (LLC). A great number of people have collaborated in AMS and ATTA experiments and in other aspects of the field on which this review is based. We acknowledge the contributions of A.M.A. Abdallah, T. Antaya, D. Anthony, K. Bailey, R. Becker, M. Bichler, T. Bigler, W. Broecker, J.Caggiano, C.Y. Chen, D. Cole, T.P. O'Connor, B. Davids, Y. Dawood, L. DeWayne Cecil, X. Du, Z. El Alfy, B. El Kaliouby, Y. El Masri, M. Fauerbach, R. Golser, R. Harkewicz, A. Heinz, M. Hellstrom, D. Henderson, C.L. Jiang, B.E.

Lehmann, P. Leleux, Y.M. Li, R. Lorenzo, A. Love, D.J. Morrissey, P. Mueller, R.C. Pardo, L.J. Patterson, M. Paul, K.E. Rehm, L. Sampson, P. Schlosser, R.H. Scott, B.M. Sherrill, W.M. Smethie,Jr., M.Steiner, N.C. Sturchio, M. Sultan, R. Vondrasek, K.D.A. Wendt, G. Winkler, L. Young. Z.-T. Lu is supported by the U.S. Department of Energy, Office of Nuclear Physics under contract W-31-109-ENG-38.

**LITERATURE CITED**

1. Burbridge EM, Burbridge GR, Fowler WA, Hoyle F. *Rev. Mod. Phys.* 29: 547 (1957)
2. Lal D, Peters B. *Handbuch der Physik*, K. Sitte, ed. Vol. XLVI/2, pp. 551-612, Berlin: Springer-Verlag (1967)
3. Masarik J, Beer J. *J. Geophys. Res. – Atmos.* 104: 12099 (1999)
4. Kutschera W, *Proc. Conf. Exp. Nucl. Phys. in Europe: Facing the Next Millennium, Sevilla, Spain, 1999*, ed. B Rubio, M Lozano, W Gelletly, AIP Conf. Ser. 495: 407 (1999)
5. Firestone RB, Shirley VS. Table of Isotopes. New York: John Wiley and Sons (1996)
6. Brosi AR, Zeldes H, Ketelle BH. *Phys. Rev.* 79: 902 (1950)
7. Stoenner RW, Schaeffer OA, Katcoff S. *Science* 148/3675: 1325 (1965)
8. Loosli, HH. *Earth Planet. Sci. Lett.* 63: 51 (1983)
9. Loosli HH, Oeschger H. *Earth Planet. Sci. Lett.* 7: 67 (1969)
10. Oeschger H, et al. *Proc. Conf. Isotope Techniques in Groundwater Hydrology*, pp. 179-190. Vienna: IAEA (1974)
11. Florkowski T, *Proc. Consultant Meeting* on *Isotopes of Noble Gases as Tracers in Environmental Studies*, pp.11-77. Vienna: IAEA (1989)
12. Lehmann BE, Davis SN, Fabryka-Martin JT. *Water Res. Res.* 29/7: 2027 (1993)
13. Reynolds JH. *Phys. Rev.* 79: 886 (1950)
14. Axelsson H, et al. *Phys. Lett.* B 210: 249 (1988)
15. Eastwood TA, Brown F, Werner RD, *Can. J. Phys.* 42: 218 (1964)
16. Baglin CM. *Nucl. Data Sheet* 69: 267 (1993)
17. Phillips FM, *Environmental Tracers in Subsurface Hydrology*, ed. P Cook, AL Herczeg, pp. 299-348. London: Kluwer Academic (2000)
18. Lehmann BE. et al. *Earth Planet Sci. Lett.* 211: 237 (2003)
19. Kutschera W, et al. *Nucl. Instrum. Methods* B92: 241 (1994)
20. Kuzminov VV, Pomanski AA. *Proc. 18th Int. Cosmic Ray Conf.,* Int. Union Pure and Applied Physics, Bangalore, India, 2: 357 (1983)
21. Collon P, et al. *Nucl. Instrum. Methods* B123: 122 (1997)
22. Masarik J. Private communication (2003).
23. Levin I, Hessheimer V, *Radiocarbon* 42/1: 69 (2000); and Levin I, private comm.
24. Collon P, et al. *Radiochimica Acta* 85: 13 (1999)
25. Zeldes H, Ketelle BH, Brosi AR. *Phys. Rev.* 79: 901 (1950)
26. Meyer RA, et al. *Phys. Rev.* C 21: 2590 (1980)
27. von Hippel F, Albright DH, Levi BG. *Sci. American* 253**:** 40 (1985)
28. Weiss W, Sartorius H, Stockurger H. *Proc. Consultant Meeting* on *Isotopes of Noble Gases as Tracers in Environmental Studies*, pp.29-72. Vienna: IAEA (1989)
29. Weiss WA, Sittkus H, Stockburger H, Sartorius H. *J. of Geophysical Res.* 88: 8574 (1983)
30. Jacob DJ, Prather MJ, Wofsy SC, McElroy MB. *J. of Geophysical Research* 92: 6614 (1987).
31. Zimmermann PH, Feichter HK, Rath HK, Crutzen PJ, Weiss W. Atmosphere Environment 23/1: 25 (1989)
32. Rozanski K, Florkowski T. *Isotope Hydrology*, Vol. II, pp. 949-961. IAEA, Vienna (1979).
33. Loosli HH. *Proc. Consultant Meeting* on *Isotopes of Noble Gases as Tracers in Environmental Studies*, pp.73-85. Vienna: IAEA (1989)
34. Smethie WM Jr., Solomon DK, Schiff SL, Mathieu GG. *J. of Hydrology,* 130: 279 (1992)

35. Ekwurzel B, et al. *Water Res. Res.* 30: 1693 (1994).
36. Loosli HH  Lehmann BE,  Smethie WMJ, *Environmental Tracers in Subsurface Hydrology*, ed. P Cook, AL Herczeg, pp. 379-397. London: Kluwer Academic (2000)
37. Smethie WM Jr, Ostlund HG, Loosli HH. *Deep Sea Res.* 33: 675 (1986).
38. Smethie WM Jr., Swift JH. *J. of Geophysical Res.* 94: 8265 (1989)

39. Stute M, Schlosser P, *Environmental Tracers in Subsurface Hydrology*, ed. Cook P, Herczeg AL, pp. 349-377. London: Kluwer Academic (2000)
40. Loosli HH, Oeschger H. *Earth Planet. Sci. Lett.* 5: 191 (1968)
41. Loosli HH, M. Heimann M, Oeschger H. *Radiocarbon* 22:461 (1980)
42. Oeschger H, Beer J, Loosli HH,  Schotterer. U. Low-level counting systems in deep underground laboratories. *in* IAEA International Conference, Vienna (1981)
43. Loosli HH, Moell M, Oeschger H,  Schotterer U. *Nucl. Instrum Meth.* B **17**:402 (1986)
44. Forster, M., K. Ramm, and P. Maier. 1992. *Proc. Consultant Meeting on Isotope Techniques in Water Resource Development*, pp. 203-214. Vienna: IAEA (1992)
45. Litherland AE, *Ann. Rev. Nucl. Part. Sci.* 30: 437 (1980)
46. Kutschera W. *Radiocarbon* 25/2: 677 (1983).
47. Elmore D, Philips FM. Science 236: 543 (1987)
48. Kutschera W, Paul M, *Ann. Rev. Nucl. Part. Sci.* 40: 411 (1990)
49. Finkel RC, Suter M. *Advances in Anal. Geochem.* 1: 1 (1993)
50. Tuniz C, Bird JR, Fink D, Herzog GF. *Accelerator Mass Spectrometry: Ultrasensitive Analysis for Global Science*, pp. 374. Boca Raton: CRC Press (1998).
51. Fifield LK. *Rep. Prog. Phys.* 62: 1223 (1999)
52. P. Collon, et al. *Earth and Planet Sci. Lett.* 182: 103(2000).
53. Paul M, et al. *Nucl. Instrum. Methods* B273: 403 (1988)
54. Betz HD. *Rev. Mod. Phys.* 44 : 465 (1972)
55. Rehm KE, Wolfs FLH. *Nucl. Instrum. Methods* A273: 262 (1988)
56. Paul M, Harrs B, Henderson D, Jiang CL, Rehm KE. *Annual Physics Division Report of Argonne National Laboratory* 97/14: 79 (1996)
57. P. Collon, et al. *Nucl. Instrum. Methods* B, to be published (2004)
58. Lu ZT, Wendt KDA. Rev. Sci. Instr. 74: 1169 (2003)
59. Letokhov VS. *Laser Photoionization Spectroscopy*. Orlando: Academic (1987).
60. Hurst GS, Payne MG. *Principles and Applications of Resonance Ionisation Spectroscopy*. New York: Adam Hilger (1988)
61. Thonnard N, Willis RD, Wright MC, Davis WA, Lehmann BE, *Nucl. Instrum. Methods* B29**:** 398 (1987)
62. Thonnard N, Payne MG, Lu Deng. *Rev. Sci. Instrum.* 65/8: 2433 (1992)
63. Thonnard N, Lehmann BE, Proc. 7$^{th}$ Int. Symp. Resonance Ionization Spectroscopy, Bernkastel-Kues, Germany, 1994, ed. HJ Kluge, JE Parks, K Wendt. *AIP Conf. Proc.* 329: 335 (1995)
64. Lehmann BE, Loosli HH, Rauber D, Thonnard N, Willis RD. *Appl. Geochem.* 6: 419 (1991)
65. Craig H, et al. *EOS* 71: 1825 (1990)
66. Fairbank WM. *Nucl. Instrum. Methods* B29: 407 (1987)
67. Fairbank WM, et al. *SPIE* 3270: 174 (1998)
68. Chen CY et al. *Science.* 286: 1139 (1999)
69. Metcalf HJ, van der Straten P. *Laser Cooling and Trapping* New York: Springer (1999).
70. Balykin VI, Minogin VG, Letokhov VS, *Rep. Prog. Phys.* 63: 1429(2000)

71. Raab EL, Prentiss M, Cable A, Chu S, Pritchard DE. *Phys. Rev. Lett.* 59: 2631 (1987)
72. Du X et al., Geophys. Res. Lett. 30: 2068 (2003)
73. Coplen TB, *Uses of environmental isotopes,* in ed. Alley WM, Regional Ground-Water Quality. New York: Van Nostrand Reinhold (1993)
74. Stute M, Schlosser P, Clark JF, Broecker WS, Science 256: 1000 (1992)
75. Beyerle U, et al. *Science* 282: 731 (1998)
76. Plummer LN, Michel RL, Thurman EM, Glynn PD, *Environmental tracers for age dating young ground water*. In ed. Alley WM, Regional Ground-Water Quality. New York: Van Nostrand Reinhold (1993)
77. Torgersen T, et al. *Water Resources Res.* 27/12: 3201 (1991)
78. Sturchio NC et al., Geophys. Res. Lett., in press (2004)
79. Broecker WS. Nat. Hist. Mag. 97: 74 (1987)
80. Broecker WS, Peng TS. *Nucl. Instrum. Methods* B172: 473 (2000)
81. Deleersnijder E., Campin JM, Delhez E. *J. Mar. Sys.* 28: 229(2001)
82. Broecker WS. *Science* 278: 1582 (1997)
83. Broecker WS, Sutherland S, Peng TS. *Science* 286: 1132 (1999)
84. Rodriguez J. *Beiträge zur Verteilung von $^{39}$Ar im Atlantik*. PhD Thesis, University of Bern Switzerland (1993)
85. Loosli HH, Applications of $^{37}$Ar, $^{39}$Ar and $^{85}$Kr in hydrology, oceanography and atmospheric studies, in *Isotopes of Noble Gases as Tracers in Environmental Studies*, pp. 73-86, IAEA, Vienna, 1992
86. Smethie WM Jr., Mathieu G. *Marine Chemistry* 18: 17 (1986)
87. Champ, MA, et al. *Marine Pollution Bulletin* 35: 203 (1998)
88. OTA. *Nuclear Wastes in the Arctic: An Analysis of Arctic and Other Regional Impacts from Soviet Nuclear Contamination*, OTA-ENV-623, Washington, DC: U.S. Government Printing Office ( September 1995)
89. Smethie WM Jr., *Arctic Nuclear Waste Assessment Program Summary FY 1995*. Office of Naval Research, Arlington, VA, pp. 192 (1996)
90. Graedel TE, Crutzen PJ. *Chemie der Atmosphäre*. Heidelberg: Spektrum Akademischer Verlag (1994)
91. Dansgard W, et al. *Nature* 364: 218 (1993)
92. Grootes PM, et al. *Nature* 366: 552 (1993)
93. Petit JR, et al. *Nature* 399: 429 (1999)
94. Keeling CD, Whorf TP, Scripps Institute of Oceanography (2003), http://cdiac.esd.ornl.gov/trends/co2/sio-mlo.htm
95. Etheridge DM, et al. *J. Geophys. Res.* 103: 15,979 (1998)
96. Johnsen SJ, et al. *Tellus* 47B: 624 (1995)
97. Thompson LG. et al. *Science* 298: 589 (2002)
98. Young L, Yang D, Dunford RW. *J. Phys.* B35: 2985 (2002)

**Table 1.** Typical times and sample sizes required for a measurement with 10% precision on a modern sample. Assuming an extraction efficiency of 50%, approximately 250 mL-STP of Ar and 50 µL-STP of Kr can be extracted from one ton of water or ice (39).

| Method | $^{39}$Ar | | | $^{81}$Kr | | | $^{85}$Kr | | |
|---|---|---|---|---|---|---|---|---|---|
| | water sample (L) | argon (mL-STP) | counting time (h) | water sample (L) | krypton (µL-STP) | counting time (h) | water sample (L) | krypton (µL-STP) | counting time (h) |
| LLC | 3,000 | 700 | 960 | | n. a. | | 400 | 20 | 72 |
| AMS | 8 | 2 | 9 | 8,000 | 400 | 9 | | n. a. | |
| ATTA | | n. a. | | 1,000 | 50 | 10 | 40 | 2 | 10 |

**FIGURE CAPTION**

**Fig. 1.** Applicable age ranges of the three noble gas radionuclides compared to those of a few other standard radionuclides.

**Fig. 2.** The atmospheric $^{85}$Kr content in the Northern Hemisphere (annual mean values) (28).

**Fig. 3.** Schematic of a shielded gas proportional counter used in LLC measurements. For $^{85}$Kr counting, the counters are filled with a P10 (10% $CH_4$, 90% Ar) + Kr mixture at pressures of 2-5 bars. For $^{39}$Ar counting, a mixture of argon with 5-10% $CH_4$ is used at pressures of 5-25 bars. The measurements are performed 35 meters below the surface in an underground laboratory in Bern. Low activity shield and an anti-coincidence arrangement further reduce the background count rate.

**Fig. 4.** Proportional counter pulse height spectrum of a $^{85}$Kr sample. The sample size is 28 μl-STP of Kr and the integrated count rate is 0.7 cpm. This corresponds to a $^{85}$Kr/Kr ratio of $1.2 \times 10^{-11}$. Counts are analyzed as function of energy above a cut off energy of about 3 keV. The high voltage is adjusted using the 8.5 keV $K_\alpha$ x-ray line induced in copper with an external $^{241}$Am calibration source. Only a fraction of the decay energy ($E_{max} = 560$ keV) is lost within the counting gas before the β-particles reach the walls of the counter.

**Fig. 5.** Schematic of the AMS experimental setup at the National Superconducting Cyclotron Laboratory, Michigan State University, used to measure $^{81}$Kr/Kr ratios. $^{81}$Kr$^{17+}$ ions are produced in the Superconducting Electron Cyclotron Resonance (SCECR) source, and are then accelerated to 45 MeV/nucleon by the cyclotron. After passing through the Be foil, the fully-stripped $^{81}$Kr$^{36+}$ ions are separated from $^{81}$Br$^{35+}$ ions using the A1200 mass spectrometer, and detected in two ΔE-E telescopes.

**Fig. 6.** Typical TOF vs. ΔE mass identification spectrum for a $^{81}$Kr measurement on one of the groundwater samples from the Great Artesian Basin in Australia. The arrow in front of the label Time-of-flight indicates the direction of increasing flight time. The encircled groups indicate the regions of the spectrum identified as $^{81}$Kr and $^{79}$Br ions. The $^{81}$Kr group contains both primary $^{81}$Kr ions from the sample material and $^{81}$Kr produced by nuclear reactions of the intense $^{81}$Br background in the Be stripper foil. The $^{79}$Br group (like all other particles in the spectrum) are reaction products of $^{81}$Br. The background of $^{81}$Kr particles were determined from a Kr samples containing no $^{81}$Kr by normalizing to the $^{79}$Br peak intensity. This leads to a $^{81}$Br-generated contribution of 12 out of the 79 observed $^{81}$Kr particles.

**Fig. 7.** Schematic view of the AMS setup at ATLAS, Argonne National Laboratory, used to measure $^{39}$Ar/Ar ratios. $^{39}$Ar$^{8+}$ ions are alternately produced in ECR I and ECR II sources, accelerated to 232 MeV by the linac, separated from their isobar $^{39}$K$^{8+}$ ions in the gas-filled spectrograph, and finally detected by an ionization chamber.

**Fig. 8.** Typical focal plane position vs. $\Delta$E signal obtained for a neutron activated argon sample using the ionization chamber (IC) coupled to the position sensitive PPAC detector. $^{39}$Ar counts are clearly separated from the intense isobaric $^{39}$K background, and from a $^{44}$Ca background peak, whose mass-to-charge ratio (44/9) happens to be close to the one of $^{39}$Ar (39/8) within 2.8‰. Between the two groups of $^{39}$Ar and $^{39}$K counts is a third peak due to pile-up induced by the intense $^{39}$K beam. The conditions improved substantially in later runs as seen in Fig. 12, where the $^{39}$K background was reduced by a factor of 130 through special measures in the ECR source (see Sec. 3.2.2).

**Fig. 9.** Schematic layout of the atom-trap beamline. Metastable krypton atoms (Kr*) are produced in the discharge. The $^{81}$Kr* atoms are transversely cooled, slowed and trapped by the laser beams shown as solid arrows. The fluorescence of individual trapped $^{81}$Kr* atoms is imaged to a detector. Total length of the apparatus is about 2.5 meters.

**Fig. 10.** Signal of a single trapped $^{81}$Kr atom. The background is due to light scattered off walls. The count-rate goes to zero between peaks because the counter is blocked during the loading of the trap. Single atom signal ~ 1600 photon counts, background ~ 340 photon counts.

**Fig. 11.** Map showing sample locations (red circles) and their $^{81}$Kr ages (in units of $10^5$ years) in relation to oasis areas (shaded green), Precambrian basement outcrops (patterned), and other regional features. Groundwater flow in Nubian Aquifer is toward northeast. For an interpretation of the deduced $^{81}$Kr ages see Sec. 4.1.2.

**Fig. 12.** Conceptual illustration of the Atlantic conveyor-belt circulation as first envisioned by W. Broecker (79). Warm shallow water is chilled in the far North Atlantic, therefore becoming more salty and sinking into the abyss. The cold and salty current flowing south near the bottom promotes a compensating northward surface layer flow of the warm, lower salinity water. The lower part of the figure illustrates the AMS detection of $^{39}$Ar from a water sample off the eastern coast of Brazil (water depth 4700 m). The few $^{39}$Ar events are cleanly separated from a huge isobaric $^{39}$K background in the gas-filled magnet (see Fig. 7), and resulted in an isotopic ratio of $^{39}$Ar/$^{40}$Ar = (2.6±0.6)×$10^{-16}$. This corresponds to 32% of the modern atmospheric ratio, and can be interpreted as an "age" of 440 years since this water parcel last interacted with the atmosphere.

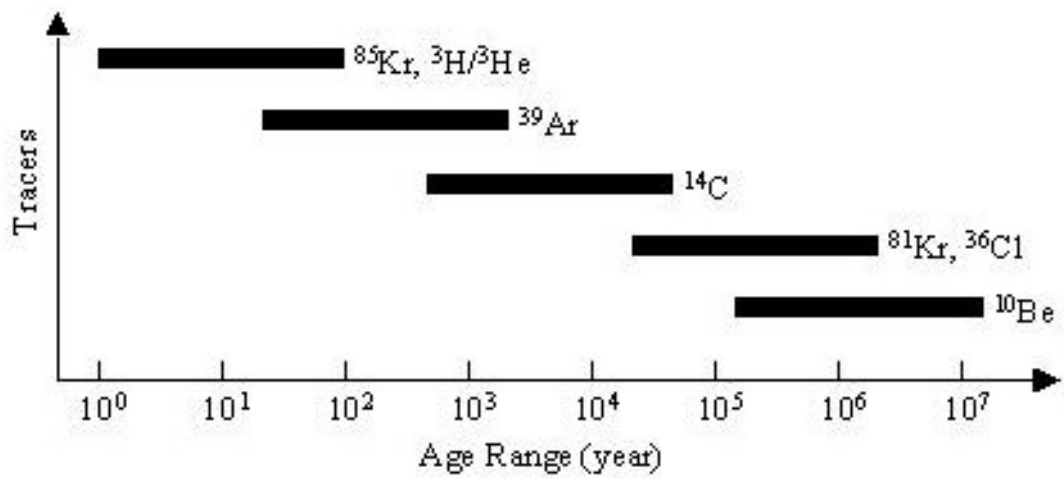

Figure 1

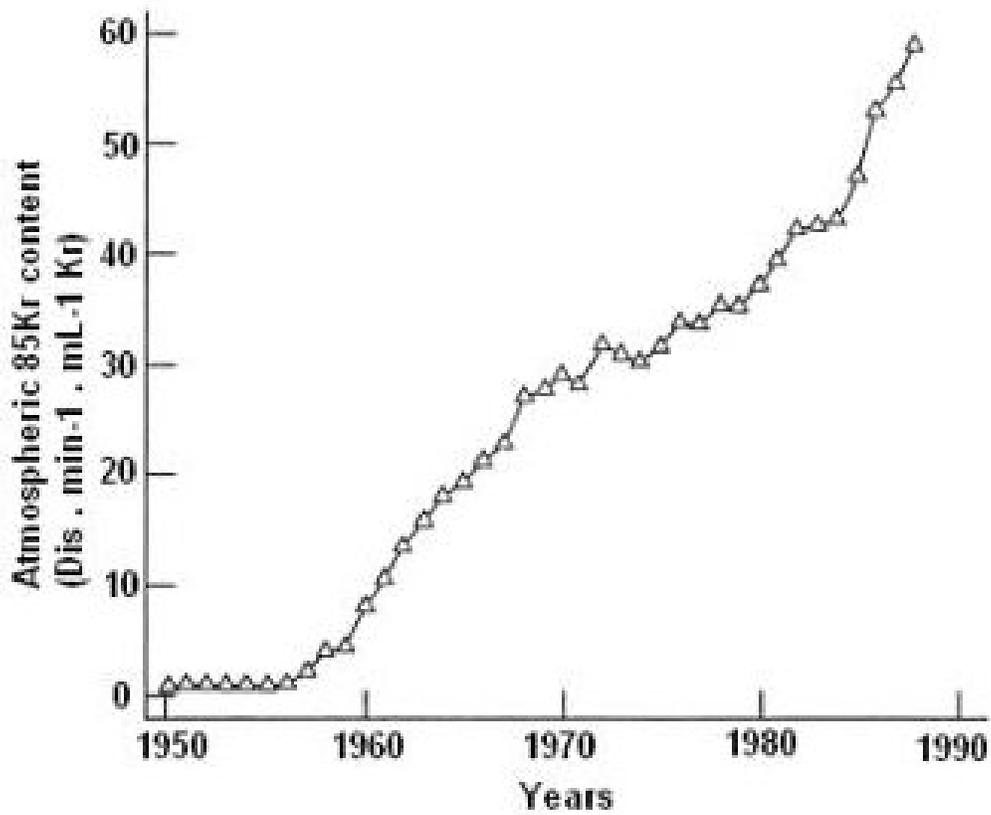

Figure 2

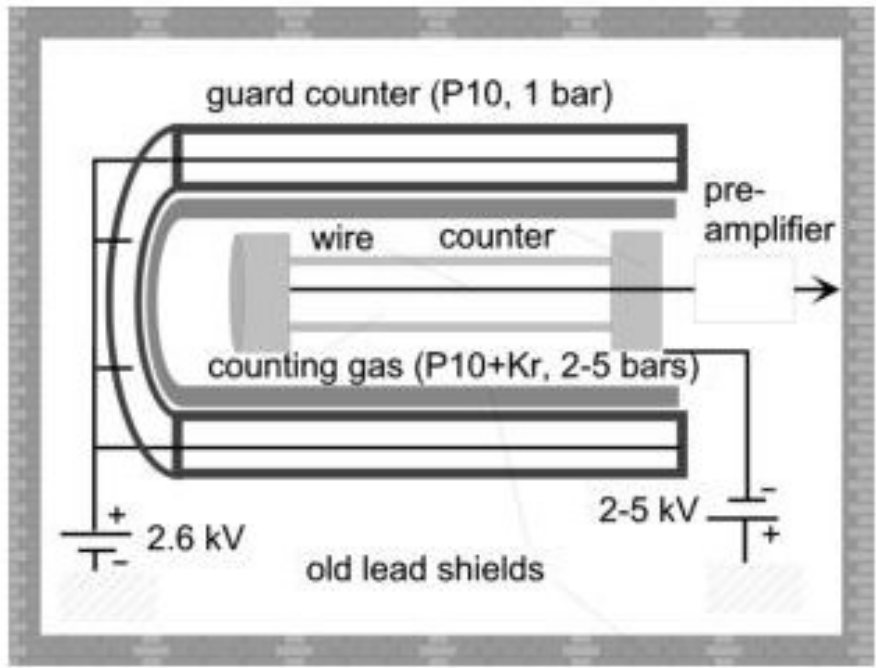

Figure 3

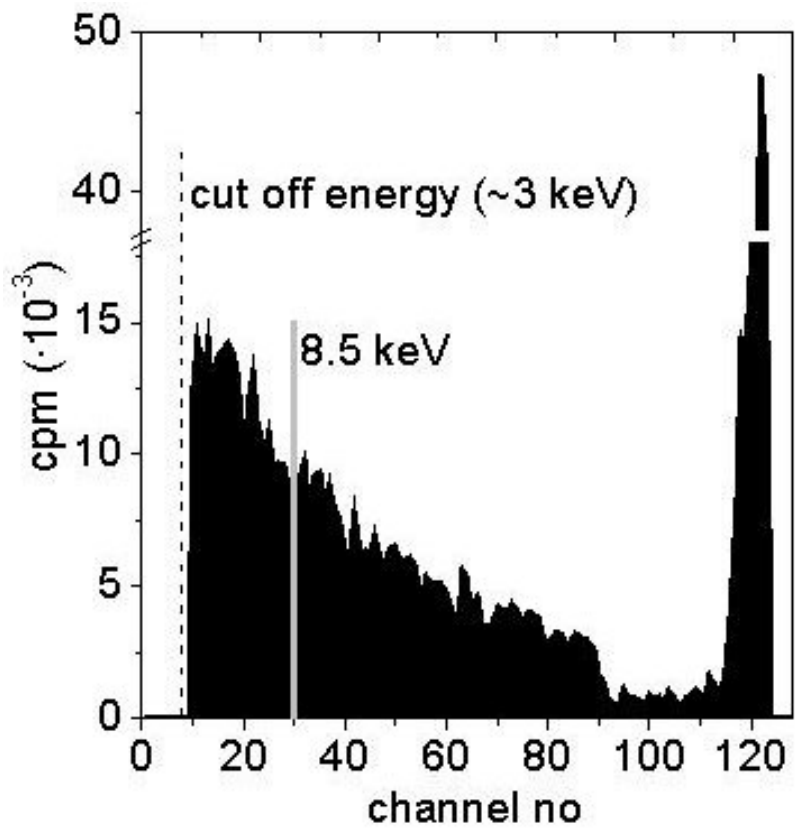

Figure 4

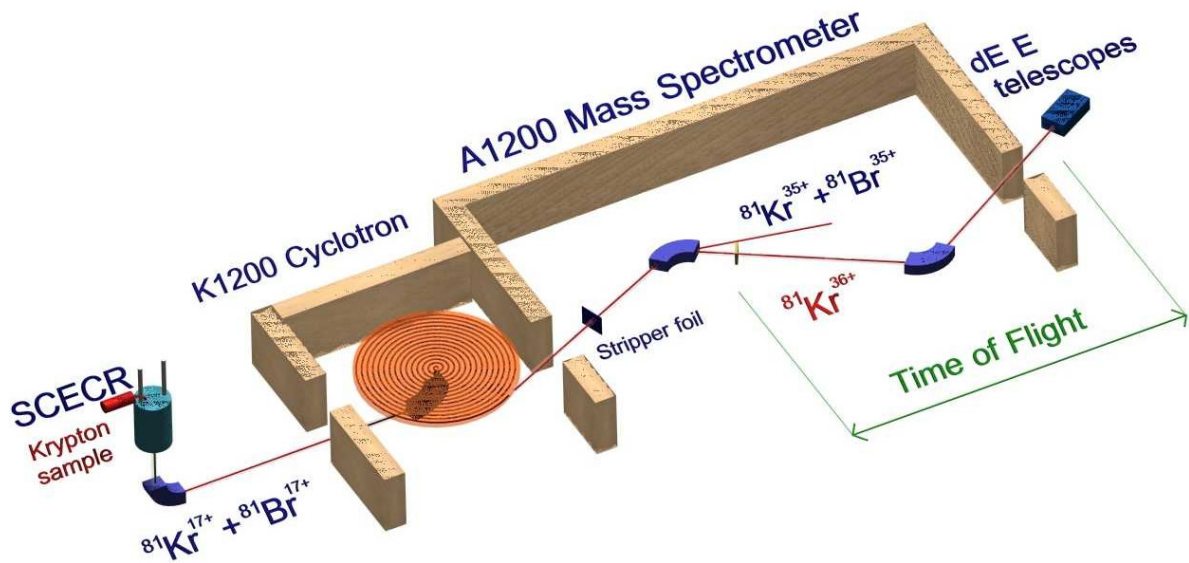

Figure 5

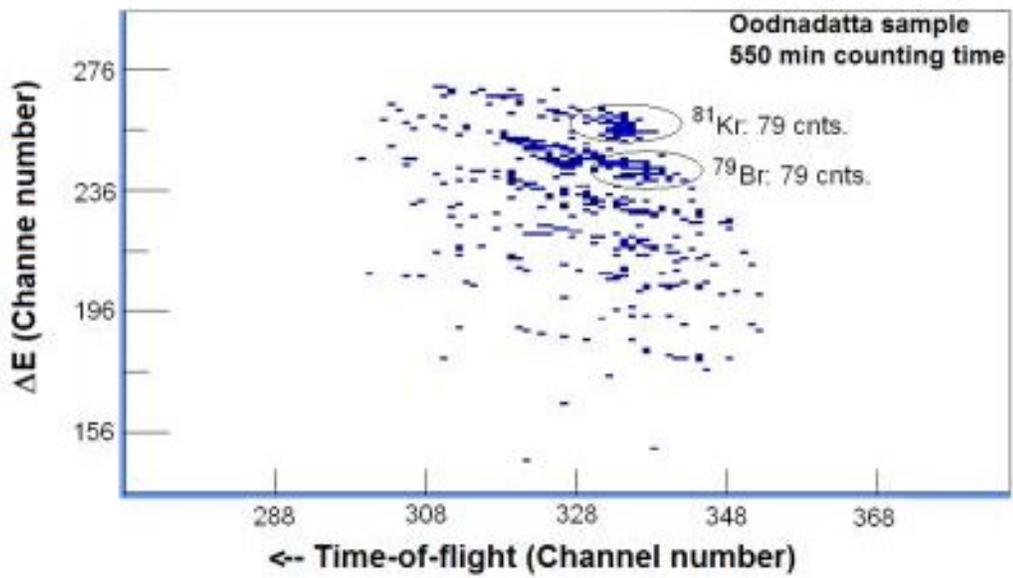

Figure 6

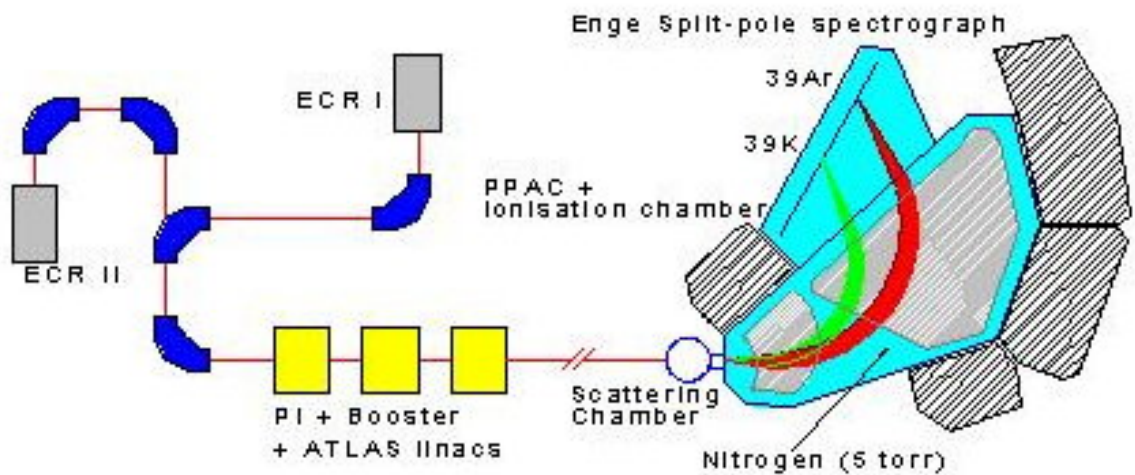

Figure 7

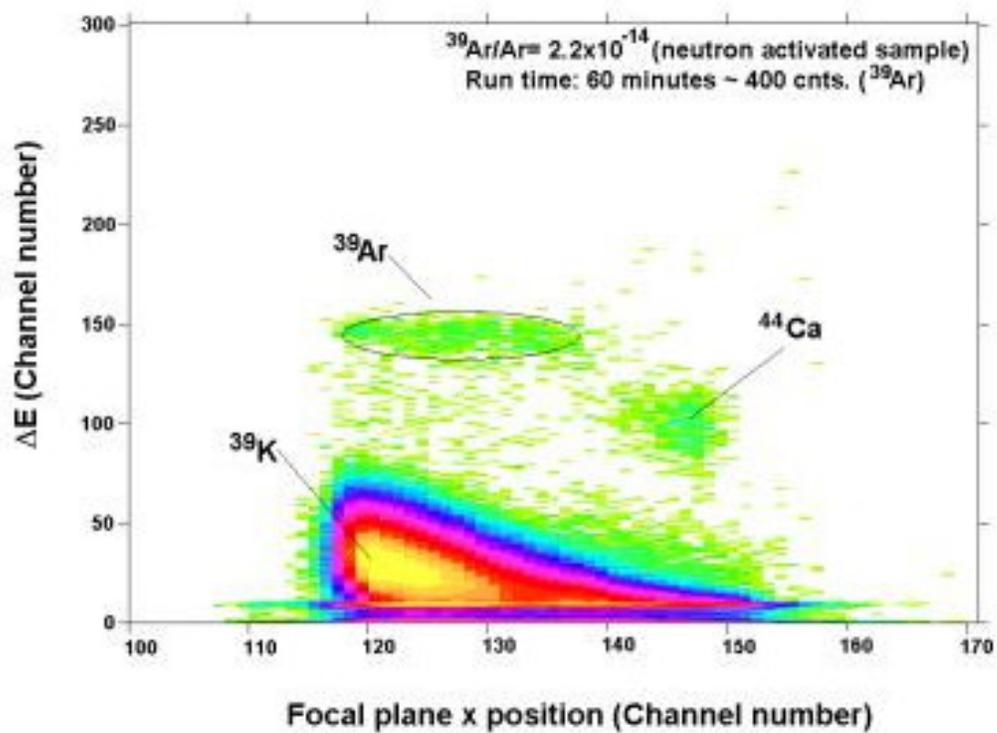

Figure 8

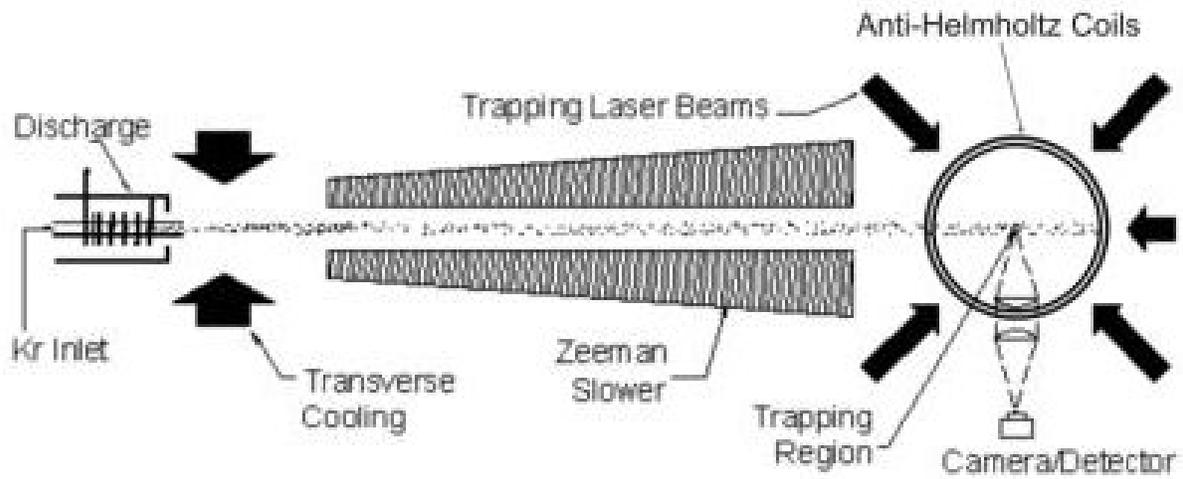

Figure 9

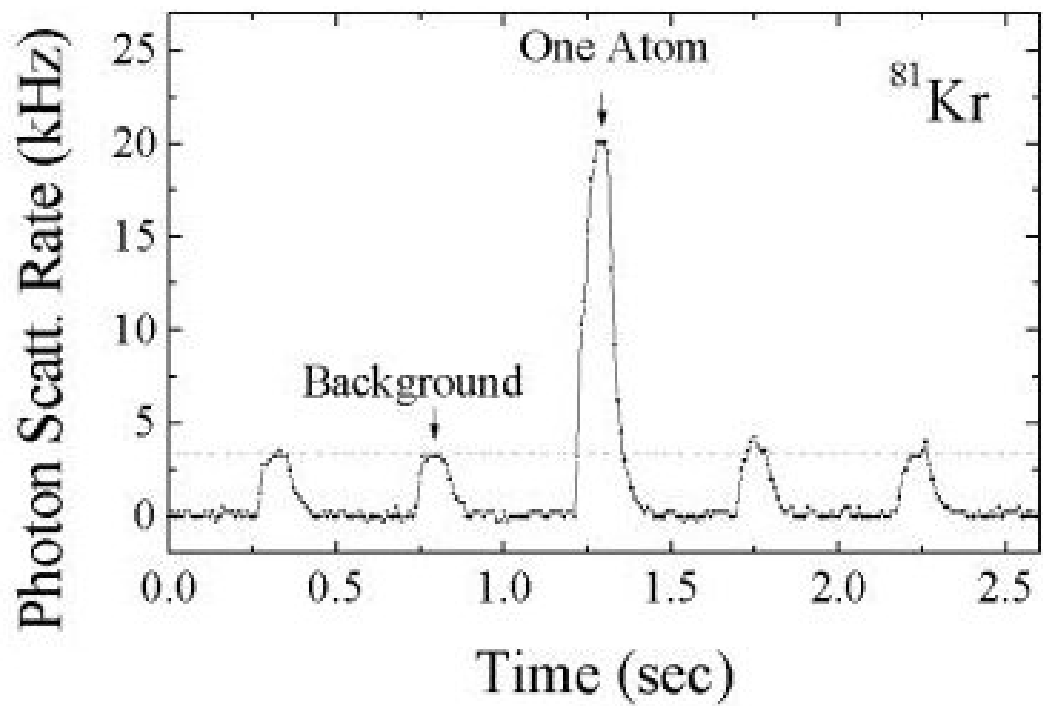

Figure 10

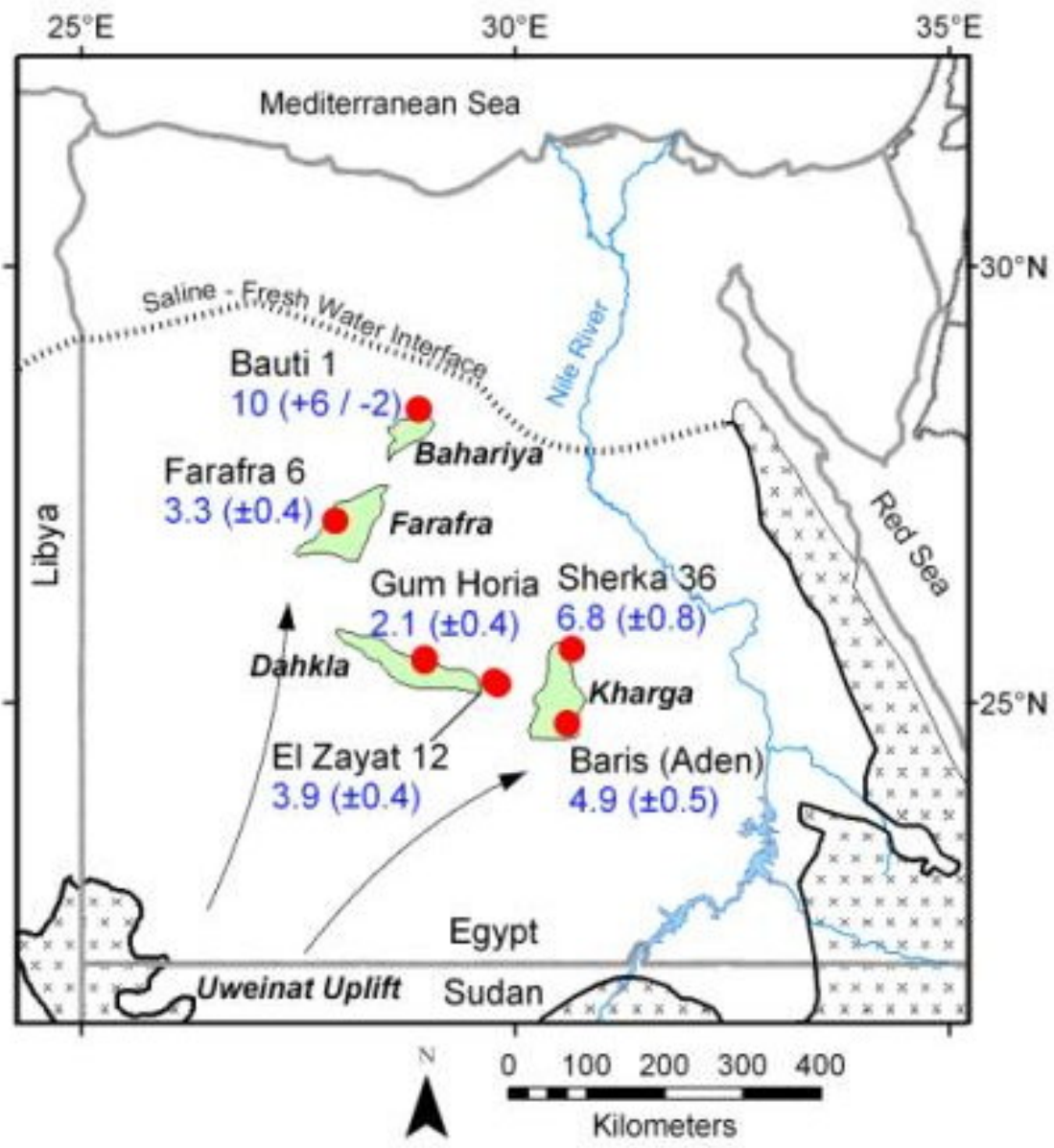

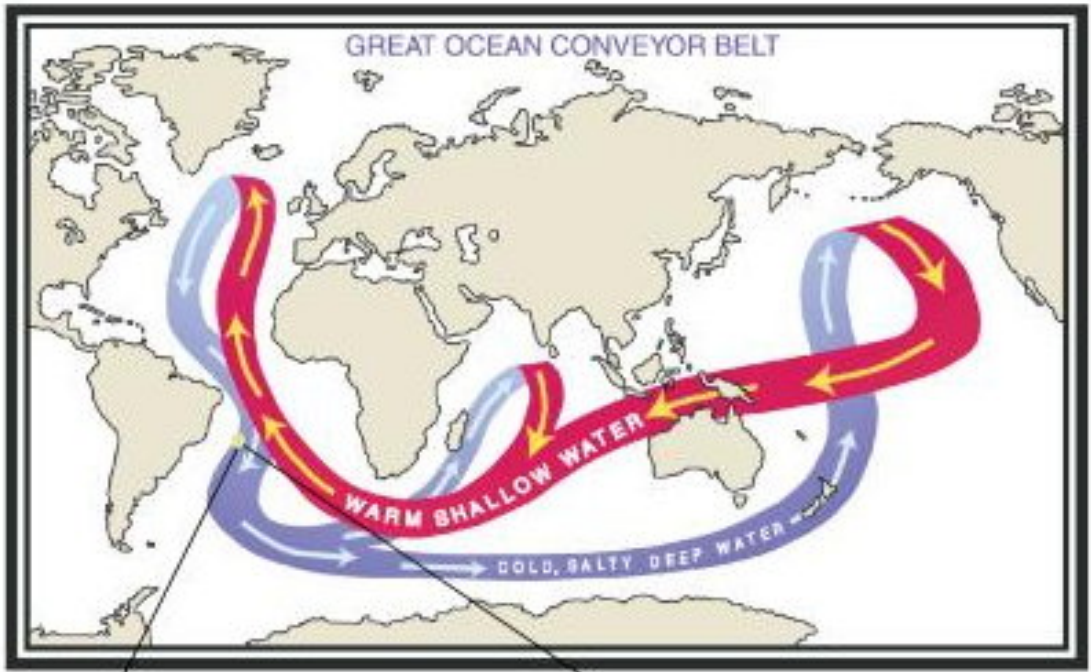

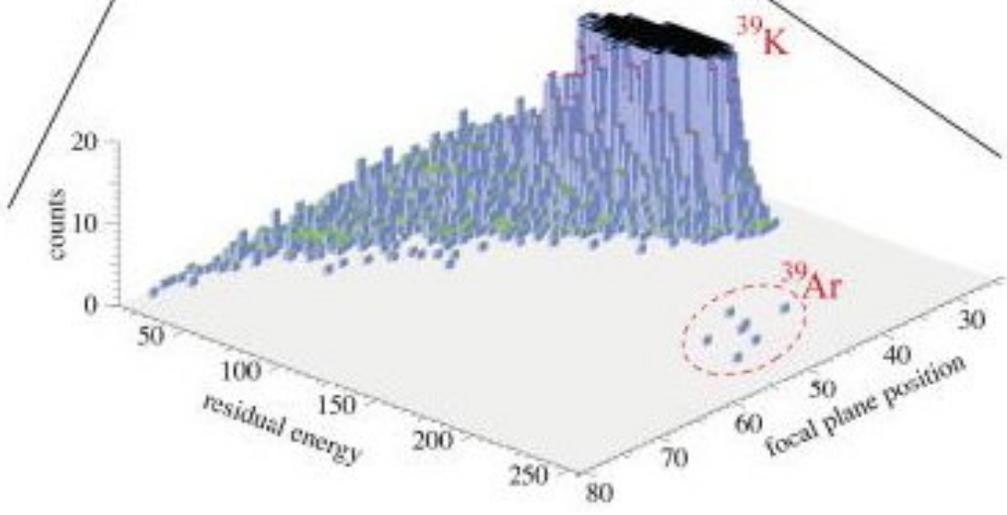

Figure 12